June 17, 2009

# On the problem of *Eigenschaften* in the Quantum and Classical Mechanics

Daniel Sepunaru

RCQCE - Research Center for Quantum Communication,

Holon Academic Institute of Technology,

52 Golomb St., Holon 58102, Israel

and

Uzi Notev

IAI

**Abstract**

We argue that in contrast to the classical physics, the measurements in the quantum mechanics should provide simultaneous information about all relevant relative amplitudes (pure states and the transitions between them) and all relevant relative phases. Simultaneity is needed since in general the measurement changes the state of the system (in quantum physics and in classical physics as well). We call that measurement procedure the holographic detection. Mathematically it is described by the set of mutually commuting self adjoint operators similar and closely related to projections. We present explicit examples and discuss general features of the corresponding experimental setup which we identify as the quantum reference frame.

> "Whence arises all that order and beauty we see in the world?
>
> Physics, beware of metaphysics."
>
> I. Newton.

**1. Introduction.**

The debates about the interconnection between the hidden laws of nature and our ability to extract the information necessary to formulate them have perhaps a history as long as study of physics itself. The content of our paper is not related to the philosophical or metaphysical aspects of those discussions. Instead, we choose certain point of view without intention to defend it or to convince the reader that it is the only possible approach. We simply present how the process of knowledge acquisition is realized within that approach. We explore the analogy with the structure of field theories (classical electrodynamics, general relativity and non-relativistic quantum mechanics) and make distinction between the unobservable kinematical quantities which characterize the physical system and the measurable variables which define its dynamics. Since the main distinction between the classical and the quantum physics is in presence of new kinematical quantities - phases, one should learn how to measure the corresponding phase differences. We demonstrate that the required measurement may be performed using special experimental arrangement which we call the quantum reference frames. The use of these reference frames allows communicating the hidden unobservable information into the instruments of the observer. Simultaneously it explains why the elementary unit of the communication is given in terms of indivisible bit.

The notion of eigenschaften operator was introduced by J.von Neumann[1] as the necessary ingredient of the theory of measurements. He suggested assigning that role to the projection operators. However, projection operators define the space of quantum mechanical states, they define the structure of that space and its orthonormal and complete basis. It is logically inconsistent to describe the alternatives by the tool that annihilates the alternative possibilities; rather it seems natural to use the operators that keep all possibilities open. Therefore, the eigenschaften operators must be closely related to projection operators but act on whole space without distortion, that is, eigenschaften operators must be unitary.

The main feature of the measurement process is that the measurement devices are macroscopic, obeying the laws of classical physics, whereas the systems under test belong to the microscopic world and behave quantum mechanically. Indeed, the measurement setup should assure that the obtained results represent the objective properties of the investigated physical system and not a free subjective imagination of the observer. Within classical physics we complete that task by introduction of reference frames such that the detector location defines the frame origin and the set of auxiliary macroscopic devices allows establishing the connection and the communication between the frames separated by the finite space-time interval (the comparison of the obtained empirical data must always be performed by the same observer). Similarly, in order to perform the measurement of the relevant quantum dynamical variable one should include in the classical setup a set of auxiliary macroscopic devices which produce the necessary beam-splitting. Then the required phase differences are measured in the usual way. This setup and recording procedure may be viewed as the general holographic detection. The organization of this paper is as follow:

Section 2 presents a discussion of the relevant kinematics of the quantum theory.

Section 3 introduces the unitary self-adjoint operators which we identify as the adequate eigenschaften operators.

Then Section 4 discusses the preferred (quantum) reference frames in close analogy to the inertial frames used in classical physics.

**2. Kinematics of quantum mechanical theory.**

We restrict ourselves to discussion of single particle states and prefer here to avoid complications introduced by special relativity. We use the orthodox kinematical approach based on mathematical framework of metric Hilbert spaces. That means that we assume that there exists at least one self-adjoint operator which generates that space. That operator is supposed to describe the dynamics of a single particle, completely isolated from the external world. All measurable quantities are also described by the self adjoint operators. In particular, the projection operators, density matrix, etc. are treated as the special kind of observables, whereas the fundamental quantity associated with the state of the physical system is a wave function. In contrast to operators which geometrically are the transformations of the given vector space, the wave functions are the vectors that form that space, are unobservable and directly immeasurable in principle.

The transition from the sterile situation of a single isolated particle to the real life physical system is achieved through the introduction of local interactions of that test particle with the fields generated by the rest of the external world. The interactions are introduced using the principle of local gauge invariance and the required complexity emerged from the statistical nature of the environment. Notice that the described approach is identical to the conventional one, established during the centuries of development of the classical physics. Only the definition of the (fundamental) interactions is connected with the new physics, since now we have to deal with the matter waves.

The fundamental property of the quantum mechanical states is expressed in terms of linear superposition principle: if $|\Psi_1>$ and $|\Psi_2>$ are two different states of the system, then

$$|\Psi> = a|\Psi_1> + b|\Psi_2> \tag{1}$$

is also a state of the system. Equivalently, we may write that algebraically:

$$|\Psi> = a\begin{pmatrix}|\Psi_1>\\0\end{pmatrix} + b\begin{pmatrix}0\\|\Psi_2>\end{pmatrix} \tag{2}$$

or

$$|\Psi> = \begin{pmatrix}a|\Psi_1>\\b|\Psi_2>\end{pmatrix}; \quad <\Psi_1|\Psi_1> = <\Psi_2|\Psi_2> = 1; \quad <\Psi_1|\Psi_2> = 0 \tag{3}$$

However, that innocent-looking mathematical expressions lead to a dramatic change in the physics of the described system, since the presence of the second orthogonal component is the necessary and sufficient condition that now the above function describes the <u>extended</u> object:

Theorem[2]: if $\hat{A}^+ = \hat{A}$ and $<\Psi_1|\Psi_2> = 0$; $<\Psi_1|\Psi_1> = <\Psi_2|\Psi_2> = 1$;

we can always decompose

$$\hat{A}|\Psi_1> = a|\Psi_1> + b|\Psi_2> \tag{4}$$

$$a = <\Psi_1|\hat{A}|\Psi_1> = <\hat{A}> \equiv \overline{A} \tag{5}$$

$$|b|^2 = bb^* = <\Psi_1|(\hat{A}^2 - a^2)|\Psi_1> \equiv (\Delta A)^2. \tag{6}$$

Proof:

$$<\Psi_1|(\hat{A}^2 - a^2)|\Psi_1> = |<\Psi_1|\hat{A}^2|\Psi_1> - a^2$$

$$= ((a<\Psi_1| + b^*<\Psi_2|)(a|\Psi_1> + b|\Psi_2>) - a^2 = bb^*$$

Therefore, what we need to reconstruct in the properly performed quantum mechanical measurement is a picture. Since the equations of motion are intrinsically complex, the quantum mechanical system must be described at least by the two component state function due to Euler relation:

$$\exp(i\varphi) = \cos\varphi + i\sin\varphi.$$

In contrast with classical physics, **the quantum mechanics is the physics of the extended objects, it is the theory of matter fields**. Now, due to D. Hilbert spectral decomposition theorem [3], any $\hat{A}$, such that $\hat{A} = \hat{A}^+$ may be expressed in terms of one-dimensional projectors:

$$\hat{A} = \sum_n \lambda_n \hat{P}_n \tag{8}$$

where

$$\hat{P}_n^+ = \hat{P}_n; \qquad \hat{P}_n \hat{P}_m = \delta_{nm} \hat{P}_m; \qquad \sum \hat{P}_n = \hat{I}; \tag{9}$$

or, in Dirac notations

$$\hat{P}_n = |\varphi_n><\varphi_n|; \tag{10}$$

$\lambda_n$ are eigenvalues of the operator $\hat{A}$ and $|\varphi_n>$ are its eigenfunctions. The set of the eigenfunctions forms the complete orthonormal basis. Thus the obtained space is the metric space suitable for the physical applications. The operator (10) defines a pure state. More generally, one introduces the density matrix

$$\hat{\rho} = |\varphi\rangle\langle\varphi|$$

$$\rho_{ij} = \langle\varphi_i|\hat{\rho}|\varphi_j\rangle = \langle\varphi_i|\varphi\rangle\langle\varphi|\varphi_j\rangle \quad (11)$$

$$\rho_{ij} = \langle\varphi_i|\varphi\rangle\langle\varphi_j|\varphi\rangle^*$$

or

$$\hat{\rho} = \sum_n w_n |\varphi_n\rangle\langle\varphi_n| \quad (12)$$

We may try to use the linear algebra machinery in order to clarify the difference between the one- and multicomponent states. In terms of Heisenberg- Schrödinger notations we may write (we consider a two-component case for simplicity only, the generalization to the non-generate finite dimension case is straightforward):

$$\hat{P}_1 = \begin{pmatrix} 1 & 0 \\ 0 & 0 \end{pmatrix} ; \quad \hat{P}_2 = \begin{pmatrix} 0 & 0 \\ 0 & 1 \end{pmatrix} \quad (13)$$

$$\hat{P}_1 + \hat{P}_2 = \hat{I}$$

Now consider the two component wave function. Then

$$|\Psi\rangle = a\begin{pmatrix} 1 \\ 0 \end{pmatrix} + b\begin{pmatrix} 0 \\ 1 \end{pmatrix} = \begin{pmatrix} a \\ b \end{pmatrix} \quad (14)$$

$$aa*+bb* = 1$$

and the corresponding density matrix[4]

$$\hat{\rho} = \begin{pmatrix} aa* & ab* \\ ba* & bb* \end{pmatrix} \quad (15)$$

may be obtained using the Kronecker product multiplication[5]:

$$\hat{\rho} = \begin{pmatrix} a \\ b \end{pmatrix} \otimes (a*, \; b*) \quad (16)$$

However, equation (15) still describes a pure state, since

$$\hat{\rho} = \hat{\rho}^+ ; \quad \hat{\rho}^2 = \hat{\rho} ; \quad Tr\hat{\rho} = 1 \quad (17)$$

Let us introduce a notation

$$\tilde{\hat{\rho}} = \begin{pmatrix} aa* & 0 \\ 0 & 0 \end{pmatrix} + \begin{pmatrix} 0 & 0 \\ 0 & bb* \end{pmatrix} \quad (18)$$

Then

$$\hat{\rho} = \hat{\tilde{\rho}} + \begin{pmatrix} 0 & ab* \\ ba* & 0 \end{pmatrix} \quad (19)$$

Obviously,

$$\hat{\tilde{\rho}} = \hat{\tilde{\rho}}^+ \ ; \ Tr\hat{\tilde{\rho}} = 1 \quad (20)$$

But

$$\hat{\tilde{\rho}}^2 \neq \hat{\tilde{\rho}} \quad \text{if } a \cdot b \neq 0 \quad (21)$$

$\tilde{\rho}$ is a mixture[7] of two single particle pure states $\begin{pmatrix} a \\ 0 \end{pmatrix}$ and $\begin{pmatrix} 0 \\ b \end{pmatrix}$. Clearly, it can not be treated as related to the two-particle state. Purpose of this example is to demonstrate that one should be careful with the density matrix techniques applications. Without reference whether the statistical interpretation of quantum theory is meaningful or not, $\tilde{\rho}$ is simply not a quantum mechanical operator at all, since it violates the most fundamental principle of quantum mechanics – linear superposition of states[8]. In addition, since

$$\hat{\tilde{\rho}}^2 \neq \hat{\tilde{\rho}}$$

$$Tr\{(\hat{\tilde{\rho}})^2\} \neq Tr\{\hat{\tilde{\rho}}\} = 1 \quad (22)$$

then the dispersion

$$(\Delta\hat{\tilde{\rho}})^2 \equiv Tr\{(\hat{\tilde{\rho}})^2\} - (Tr\{\hat{\tilde{\rho}}\})^2 = Tr\{(\hat{\tilde{\rho}})^2\} - 1 \neq 0 \quad (23)$$

and it is not classical physics operator either. In contrast, the operator $\hat{\rho}$ (eq. (15)) preserves its clear geometrical meaning of a one-dimensional projector (dispersion free). If one starts with a well defined reference frame, the complete set of those projectors allows performing the rotation to the new axes of that new reference frame. However, that set does not allow extracting the information about the dispersions contained in the measurements of the transition amplitudes. In the next section we will discuss the self-adjoint operators that allow doing that.

## 2. Eigenschaften Operators.

From the logical point of view it is natural to expect that the projection operators do not provide the adequate tool to obtain information about all possible alternatives, since they destroy the orthogonal subspace of the Hilbert space. The true eigenschaften operator must be unitary. Together with the requirement of being observable ($\hat{H}^+ = \hat{H}$) that leads to the following statement:

Theorem.

If $\hat{H}^+ = \hat{H}^{-1}$ (unitary) and $\hat{H}^+ = \hat{H}$ (self- adjoint),

Then
$$\hat{H}^2 = \hat{I} \ .$$
Proof: (24)

1) Suppose

$$\hat{H}^+ = \hat{H} = \hat{H}^{-1},$$

then

$$\hat{H} \bullet \hat{H} = \hat{H} \bullet \hat{H}^{-1} = \hat{I}.$$

2) Suppose

$$\hat{H}^2 = \hat{I} \text{ and } \hat{H}^+ = \hat{H}^{-1},$$

then

$$\hat{H}^+ = \hat{H} \ .$$

From $\hat{H}^2 = \hat{I}$ we have

$$(\hat{H} - \hat{I}) \bullet (\hat{H} + \hat{I}) = 0 \qquad (25)$$

Let us consider first the two-dimensional case. From the Eq.(25)

$$\lambda_1 = 1 \ ; \ \lambda_2 = -1 \qquad (26)$$

and due to spectral composition theorem, we have

$$\hat{H}_2 = \hat{P}_1 - \hat{P}_2 \ . \qquad (27)$$

Since
$$\hat{P}_1 + \hat{P}_2 = \hat{I},  \tag{28}$$

finally we obtain

$$\hat{P}_1 = \frac{\hat{I} + \hat{H}_2}{2}$$
$$\hat{P}_2 = \frac{\hat{I} - \hat{H}_2}{2} \tag{29}$$

Now in terms of matrix mechanics we have

$$\hat{H}_2 | \Psi_1 > = \alpha_1 | \Psi_1 > + \beta e^{i\Delta\varphi} | \Psi_2 > \equiv | \Psi_3 >$$
$$\hat{H}_2 | \Psi_2 > = \beta e^{-i\Delta\varphi} | \Psi_1 > + \alpha_2 | \Psi_2 > \equiv | \Psi_4 > \tag{30}$$

with

$$< \Psi_1 | \Psi_1 > = < \Psi_2 | \Psi_2 > = < \Psi_3 | \Psi_3 > = < \Psi_4 | \Psi_4 > = 1$$
$$< \Psi_1 | \Psi_2 > = < \Psi_3 | \Psi_4 > = 0 \tag{31}$$

Then
$$\beta \cdot (\alpha_1 + \alpha_2) = 0$$
$$\alpha_1^2 + \beta^2 = 1 \tag{32}$$
$$\alpha_2^2 + \beta^2 = 1$$

Since here we discuss the measurements of relevant parameters of quantum mechanical systems with non-vanishing dispersion, we will consider only $\beta \neq 0$ case. Then

$$\alpha_1 = -\alpha_2 \equiv \alpha \tag{33}$$

or

$$Tr(\hat{H}_2) = 0 \tag{34}$$

Using the relations (32) we obtain the following most general solution

$$\hat{H}_2 = \begin{pmatrix} \cos\gamma & e^{i\cdot\Delta\varphi} \cdot \sin\gamma \\ e^{-i\cdot\Delta\varphi} \cdot \sin\gamma & -\cos\gamma \end{pmatrix} \tag{35}$$

In particular, for $\Delta\varphi = 0$ and $\gamma = 45°$ we obtain the Hadamard matrix of lowest order (N=2)

$$\hat{H}_2 = \frac{1}{\sqrt{2}} \begin{pmatrix} 1 & 1 \\ 1 & -1 \end{pmatrix} \tag{36}$$

well known in image processing applications.

The matrix elements

$$(\hat{H}_2)_{11} = <\Psi_1|\hat{H}_2|\Psi_1> = - <\Psi_2|\hat{H}_2|\Psi_2> = \alpha \tag{37}$$

and

$$(\hat{H}_2)_{12} = <\Psi_1|\hat{H}_2|\Psi_2> = (<\Psi_2|\hat{H}_2|\Psi_1>)^* = \beta e^{i\cdot\Delta\varphi} \tag{38}$$

are all we need to know about the quantum state. Both are measurable, $(\hat{H}_2)_{11}$ define the spectrum and $(\hat{H}_2)_{12}$ define the dispersion. The basis introduced above $|\Psi_3>$ and $|\Psi_4>$ is distinguished by the fact that it allows measurement both of them simultaneously. Perhaps the example of two level systems makes that even clearer:

$$\begin{aligned}\hat{H}_2(e^{-i\cdot\omega_1 t}|\Psi_1>) &= \alpha_1 e^{-i\cdot\omega_1 t}|\Psi_1> + \beta e^{-i\cdot\omega_2 t}|\Psi_2> \\ \hat{H}_2(e^{-i\cdot\omega_2 t}|\Psi_2>) &= \beta e^{-i\cdot\omega_1 t}|\Psi_1> - \alpha e^{-i\cdot\omega_2 t}|\Psi_2>\end{aligned} \tag{39}$$

Then dropping the overall phase factor, we obtain

$$\begin{aligned}\hat{H}_2|\Psi_1> &= \alpha|\Psi_1> + \beta e^{+i(\omega_1-\omega_2)t}|\Psi_2> \\ \hat{H}_2|\Psi_2> &= \beta e^{-i(\omega_1-\omega_2)t}|\Psi_1> - \alpha|\Psi_2>\end{aligned} \tag{40}$$

Let us consider now the three component case (analog to three level quantum mechanical systems).

We prefer to discuss explicitly the three component case and the four component case, rather than development of the general n-dimensional situation, which follows straightforward from the obtained results.

We have

$$\begin{aligned}\hat{H}_3|\Psi_1> &= \alpha_1|\Psi_1> + \beta e^{i\Delta\varphi_1}|\Psi_2> + \gamma e^{i\Delta\varphi_2}|\Psi_3> \equiv |\Psi_4> \\ \hat{H}_3|\Psi_2> &= \beta e^{-i\Delta\varphi_1}|\Psi_1> + \alpha_2|\Psi_2> + \mu e^{i\Delta\varphi_3}|\Psi_3> \equiv |\Psi_5> \\ \hat{H}_3|\Psi_3> &= \gamma e^{-i\Delta\varphi_2}|\Psi_1> + \mu e^{-i\Delta\varphi_3}|\Psi_2> + \alpha_3|\Psi_3> \equiv |\Psi_6>\end{aligned} \tag{41}$$

with

$$<\Psi_1|\Psi_1> = <\Psi_2|\Psi_2> = <\Psi_3|\Psi_3> = 1$$
$$<\Psi_4|\Psi_4> = <\Psi_5|\Psi_5> = <\Psi_6|\Psi_6> = 1$$
$$<\Psi_1|\Psi_2> = <\Psi_1|\Psi_3> = <\Psi_2|\Psi_3> = 0$$
$$<\Psi_4|\Psi_5> = <\Psi_4|\Psi_6> = <\Psi_5|\Psi_6> = 0 \tag{42}$$

Then the matrix elements of $\hat{H}_3$ are connected by the following relations

$$\begin{aligned}
(Tr(\hat{H}_3))^2 &= (\alpha_1 + \alpha_2 + \alpha_3)^2 = 1 \\
Tr(\hat{H}_3) &= \pm 1 \\
\beta^2 &= (1 \mp \alpha_1)(1 \mp \alpha_2) \\
\gamma^2 &= (1 \mp \alpha_1)(1 \mp \alpha_3) \\
\mu^2 &= (1 \mp \alpha_2)(1 \mp \alpha_3) \\
\Delta\varphi_3 &= \Delta\varphi_2 - \Delta\varphi_1
\end{aligned} \tag{43}$$

Let us establish the connection between the eigenschaften and the projection operators here. Consider the one-dimension projection operators

$$\hat{P}_1 = \begin{pmatrix} 1 & 0 & 0 \\ 0 & 0 & 0 \\ 0 & 0 & 0 \end{pmatrix}; \hat{P}_2 = \begin{pmatrix} 0 & 0 & 0 \\ 0 & 1 & 0 \\ 0 & 0 & 0 \end{pmatrix}; \hat{P}_3 = \begin{pmatrix} 0 & 0 & 0 \\ 0 & 0 & 0 \\ 0 & 0 & 1 \end{pmatrix};$$
$$\hat{P}_1 + \hat{P}_2 + \hat{P}_3 = \hat{I} \tag{44}$$

The most general one-dimensional projector again may be written in the form

$$\hat{\rho} = \begin{pmatrix} a \\ b \\ c \end{pmatrix} \otimes (a*, \ b*, \ c*)$$
$$\hat{\rho}^+ = \hat{\rho}; \hat{\rho}^2 = \hat{\rho}; Tr\hat{\rho} = aa^* + bb^* + cc^* = 1 \tag{45}$$

Then using the spectral decomposition

$$\hat{H}_3 = \lambda_1 \hat{P}_1 + \lambda_2 \hat{P}_2 + \lambda_3 \hat{P}_3$$
$$\hat{H}_3^2 = \hat{I} \tag{46}$$

we have

$$\begin{aligned}
\hat{H}_3^{(1)} &= -\hat{P}_1 + \hat{P}_2 + \hat{P}_3 \\
\hat{H}_3^{(2)} &= \hat{P}_1 - \hat{P}_2 + \hat{P}_3 \\
\hat{H}_3^{(3)} &= \hat{P}_1 + \hat{P}_2 - \hat{P}_3
\end{aligned} \tag{47}$$

Thus we obtain

$$\hat{P}_1 = \frac{\hat{I} - \hat{H}_3^{(1)}}{2}$$

$$\hat{P}_2 = \frac{\hat{I} - \hat{H}_3^{(2)}}{2} \tag{48}$$

$$\hat{P}_3 = \frac{\hat{I} - \hat{H}_3^{(3)}}{2}$$

However, only two of them are linearly independent

$$\hat{H}_3^{(1)} + \hat{H}_3^{(2)} + \hat{H}_3^{(3)} = \hat{I} \tag{49}$$

and form the following commutative algebra

$$\hat{H}_3^{(1)} \bullet \hat{H}_3^{(2)} = -\hat{H}_3^{(3)}$$
$$\left[\hat{H}_3^{(i)}, \hat{H}_3^{(j)}\right] = 0; i, j = 1, 2, 3 \tag{50}$$

We conclude with demonstration of four component case. The $\hat{H}_4$ operators ($\hat{H}_4^+ = \hat{H}_4$ and $\hat{H}_4^2 = \hat{I}$) have a form

$$\hat{H}_4 = \begin{pmatrix} \alpha_1 & \beta e^{i\Delta\varphi_1} & \gamma e^{i\Delta\varphi_2} & \delta e^{i\Delta\varphi_4} \\ \beta e^{-i\Delta\varphi_1} & \alpha_2 & \mu e^{i\Delta\varphi_3} & \upsilon e^{i\Delta\varphi_5} \\ \gamma e^{-i\Delta\varphi_2} & \mu e^{-i\Delta\varphi_3} & \alpha_3 & \zeta e^{i\Delta\varphi_6} \\ \delta e^{-i\Delta\varphi_4} & \upsilon e^{-i\Delta\varphi_5} & \zeta e^{-i\Delta\varphi_6} & \alpha_4 \end{pmatrix} \tag{51}$$

$$\Delta\varphi_3 = \Delta\varphi_2 - \Delta\varphi_1$$
$$\Delta\varphi_5 = \Delta\varphi_4 - \Delta\varphi_1 \tag{52}$$
$$\Delta\varphi_6 = \Delta\varphi_4 - \Delta\varphi_2$$

Now we have

$$Tr(\hat{H}_4) = -2, 0, 2 \tag{53}$$

If $Tr(\hat{H}_4) = \pm 2$, the transition amplitudes (dispersions) are related to the spectrum through the following equations:

$$\beta^2 = (1 \mp \alpha_1)(1 \mp \alpha_2)$$
$$\gamma^2 = (1 \mp \alpha_1)(1 \mp \alpha_3)$$
$$\delta^2 = (1 \mp \alpha_1)(1 \mp \alpha_4)$$
$$\mu^2 = (1 \mp \alpha_2)(1 \mp \alpha_3) \tag{54}$$
$$\upsilon^2 = (1 \mp \alpha_2)(1 \mp \alpha_4)$$
$$\zeta^2 = (1 \mp \alpha_3)(1 \mp \alpha_4)$$

Notice that these are universally valid relations and thus they are subject of direct experimental verification.

Similarly to above, we may establish relations between the eigenschaften and the projection operators. For example, for the $Tr(\hat{H}_4) = 2$ we obtain

$$\begin{aligned}
\hat{I} &= \hat{P}_1 + \hat{P}_2 + \hat{P}_3 + \hat{P}_4 \\
\hat{H}_4^{(1)} &= -\hat{P}_1 + \hat{P}_2 + \hat{P}_3 + \hat{P}_4 = \hat{I} - 2\hat{P}_1 \\
\hat{H}_4^{(2)} &= \hat{P}_1 - \hat{P}_2 + \hat{P}_3 + \hat{P}_4 = \hat{I} - 2\hat{P}_2 \\
\hat{H}_4^{(3)} &= \hat{P}_1 + \hat{P}_2 - \hat{P}_3 + \hat{P}_4 = \hat{I} - 2\hat{P}_3 \\
\hat{H}_4^{(4)} &= \hat{P}_1 + \hat{P}_2 + \hat{P}_3 - \hat{P}_4 = \hat{I} - 2\hat{P}_4
\end{aligned} \qquad (55)$$

Again, we have

$$\frac{1}{2}\sum_{i=1}^{4} \hat{H}_4^{(i)} = \hat{I} \qquad (56)$$

and

$$\begin{aligned}
\hat{H}_4^{(1)} \bullet \hat{H}_4^{(2)} &= \hat{H}_4^{(1)} + \hat{H}_4^{(2)} - \hat{H}_4^{(3)} - \hat{H}_4^{(4)} \\
\left[\hat{H}_4^{(i)}, \hat{H}_4^{(j)}\right] &= 0; i, j = 1,2,3
\end{aligned} \qquad (57)$$

and so on.

For the case $Tr(\hat{H}_4) = 0$, we may write

$$\begin{aligned}
\hat{H}_4^{(1)} &= I \otimes \hat{H}_2 \\
\hat{H}_4^{(2)} &= \hat{H}_2 \otimes I \\
\hat{H}_4^{(3)} &= \hat{H}_2 \otimes \hat{H}_2
\end{aligned} \qquad (58)$$

since

$$Tr(A \otimes B) = TrA \cdot TrB \qquad (59)$$

Then we have

$$\begin{aligned}
\hat{I} &= \hat{P}_1 + \hat{P}_2 + \hat{P}_3 + \hat{P}_4 \\
\hat{H}_4^{(1)} &= \hat{P}_1 - \hat{P}_2 + \hat{P}_3 - \hat{P}_4 \\
\hat{H}_4^{(2)} &= \hat{P}_1 + \hat{P}_2 - \hat{P}_3 - \hat{P}_4 \\
\hat{H}_4^{(3)} &= \hat{P}_1 - \hat{P}_2 - \hat{P}_3 + \hat{P}_4
\end{aligned} \qquad (60)$$

and

$$\hat{P}_1 = \frac{1}{4}\left[\hat{I} + \hat{H}_4^{(1)} + \hat{H}_4^{(2)} + \hat{H}_4^{(3)}\right]$$

$$\hat{P}_2 = \frac{1}{4}\left[\hat{I} - \hat{H}_4^{(1)} + \hat{H}_4^{(2)} - \hat{H}_4^{(3)}\right]$$

$$\hat{P}_3 = \frac{1}{4}\left[\hat{I} + \hat{H}_4^{(1)} - \hat{H}_4^{(2)} - \hat{H}_4^{(3)}\right] \quad (61)$$

$$\hat{P}_4 = \frac{1}{4}\left[\hat{I} - \hat{H}_4^{(1)} - \hat{H}_4^{(2)} + \hat{H}_4^{(3)}\right]$$

Again we have

$$\hat{H}_4^{(1)} \bullet \hat{H}_4^{(2)} = \hat{H}_4^{(3)}$$
$$\left[\hat{H}_4^{(i)}, \hat{H}_4^{(j)}\right] = 0; i, j = 1,2,3 \quad (62)$$

We suppose that the way to further generalization is obvious to the reader.

**4. Holographic Detection: Quantum Reference Frames.**

Perhaps nobody needs explanations of the mathematical formalism discussed in the previous section: we hope it speaks for itself. Nevertheless, we devote this section to the description of the physical "picture" behind the presented approach since that was the guideline that leads us to it.

We addressed the following questions:

1. What is the difference between "on-off" and "or-and" switches in terms of quantum mechanical self- adjoint operators (observables)?

2. How the transition amplitudes between the stationary (pure) states are incorporated naturally and symmetrically together with the amplitudes of these states?

3. Is it possible to measure $\overline{\hat{A}}$ and $\Delta\hat{A}$ simultaneously and how to arrange the required setup?

4. If it is possible, may that measurement be performed using only macroscopic devices?

5. What Heisenberg dispersion relations have to do with those measurements?

Our answer to the last question: almost nothing. It is well known [9] that the product of two noncommuting self-adjoint operators is not a self-adjoint operator and that the dispersion of their product is not a self-adjoint operator also. Therefore, there is no way to assign the physical meaning to its numerical value. The HDR has outstanding theoretical importance telling us that quantum physics is the physics of extended objects and not of the Newtonian material points. The results of measurements are "pictures" and could not be treated as an image of a single space-time point, in principle. The projection operators intensively used by J. v. Neumann in his attempt to formulate the theory of measurements obviously play a role of "on-off" switches defining the basis of the state vectors in the Hilbert space. Therefore, it is reasonable to expect that the "or-and" operators should be connected to them but different. The particular example of the suitable candidates is the Hadamard transformations [2, 10] which already find their applications in image processing and in the quantum information theory. In addition, the notions of bits and qubits appear naturally as two component wave packets. Finally, in order to provide the laboratory realization of the simultaneous measurement of the relevant amplitudes (relative generalized coordinates) and phase differences one should assure that the wave packet will arrive to each point of the detector screen.

Let us expose the content of our discussion viewed from the eyes of Schrödinger's cat totally confused by endless debates about his destiny. The usual justification of apparent uncertainty in it refers to HDR. But the empirical evidence clearly tells us that the initial assumption that the cat may be considered as quantum mechanical system containing an inherent indeterminacy which "becomes transformed into macroscopic indeterminacy" [11] is wrong. If the state of the system ("cat") is defined, one can measure its dispersion. Now, if in that given state the dispersion is not zero, we deal with an extended object and the expected result of the measurements should be represented by a picture of poor cat "mixed or smeared out in equal parts" [11]; if not, the cat was and will remain in the pure (definite) state, hopefully alive[12]. So far no problem, but N.Bohr and his surrounding scientific community, deeply affected by the "mystery" of the quantum world, claimed that the obtained experimental results have nothing to do with the original quantum mechanical object:

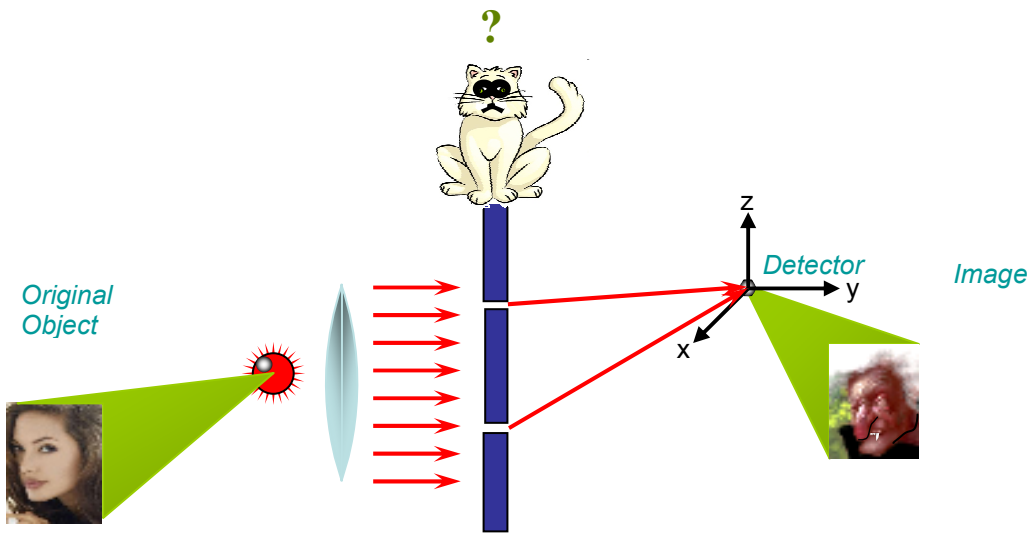

Now, let us remember what we are doing in the classical case, when only measurements of amplitudes are required. In that case, nobody doubts that the "moon is there" and in addition it is the same for all inertial reference frames:

Classical Reference Frame 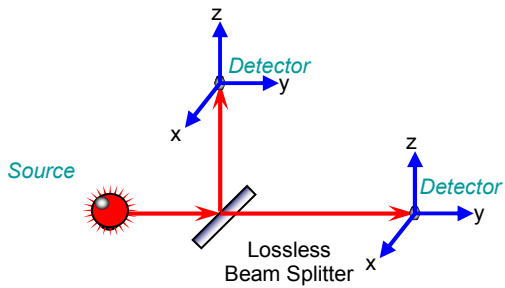 and Hanbury Brown-Twiss 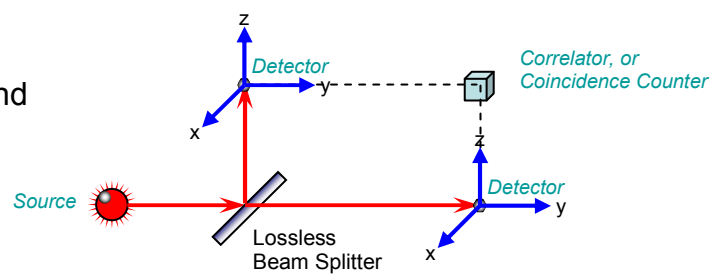

The lossless beam splitter here is the macroscopic device which participates actively in the detection procedure ($\hat{H}_2^+ = \hat{H}_2$).

In contrast, in microscopic quantum mechanical world (quantum optics) we are required to measure also the phase differences in order to obtain all existent and necessary information about the original object. This may be done using similar setup, for example,

Lloyd's Mirror          Lossless Beam Splitter

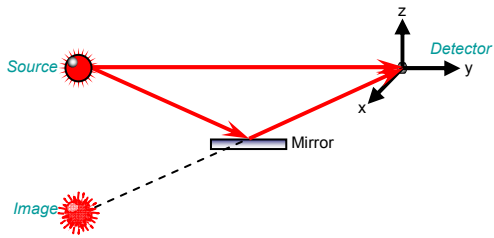 and 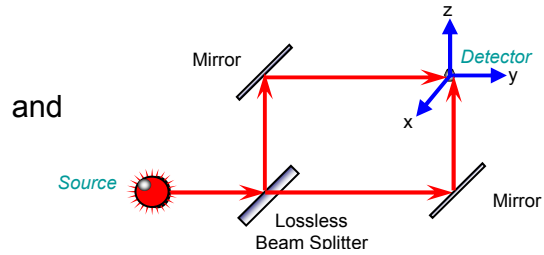

but in both cases the mirror and the lossless beam splitter participate in detection only passively; they do not cause the wave function collapse, but allow extracting the phase difference information, since the reference component of the wave packet arrives to each point of the detector screen together with the tested wave packet (within inherent dispersion of the quantum mechanical space-time continuum). Then there is no reason to expect that the obtained picture would not provide the adequate image of the original object:

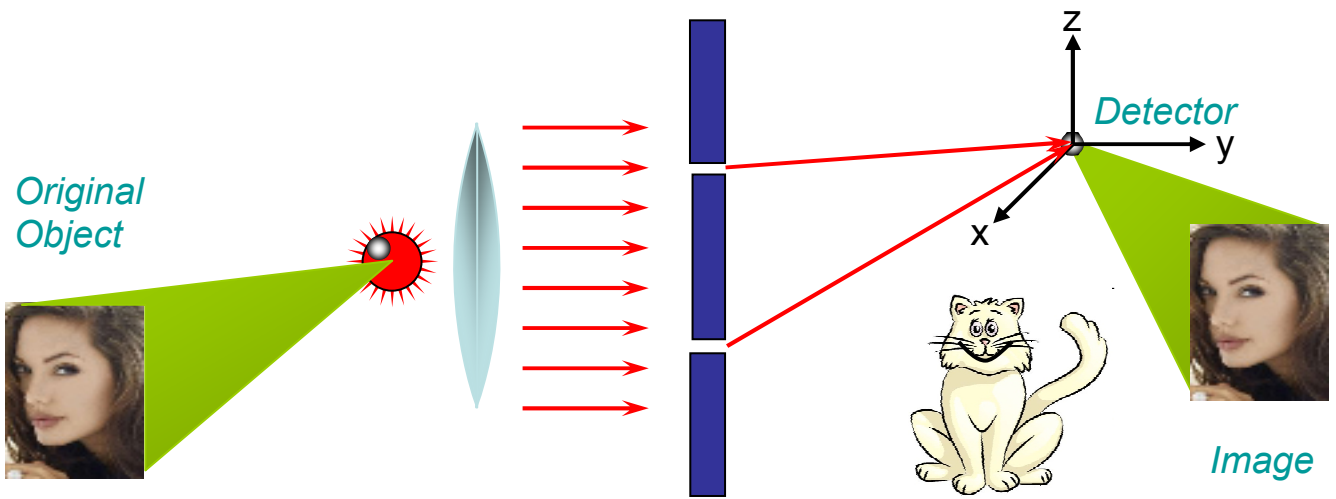

with the cat safely finding his place in the comfortable macroscopic world.